# Characterization of Recirculating Waveguide Meshes Based on an Optimization Method with a Parameter Space Reduction Technology

Ran Tao, Jifang Qiu,* Yuchen Chen, Bowen Zhang, Yan Li, Hongxiang Guo and Jian Wu

*State Key Laboratory of Information Photonics and Optical Communications, School of Electronic Engineering, Beijing University of Posts and Telecommunications, Beijing 100876, China*

Fabrication imperfections must be considered during configuration to ensure that the setup is suitable for the actual fabricated programmable photonic integrated circuits (PPICs). Therefore, characterization of imperfections is crucial but difficult, especially for PPICs made from recirculating waveguide meshes. The flexibility required by these meshes demands a more complex topology and compact TBU structure, complicating the characterization. In this paper, we propose a characterization method applicable to recirculating waveguide meshes based on an optimization approach, along with a step-by-step procedure to reduce the parameter space of optimization, allowing for characterizing imperfect parameters of each individual component within the waveguide mesh. To the best of our knowledge, this method can greatly broaden the range of characterized parameters compared to currently reported methods. In order to verify the effectiveness of our method, we used the characterized parameters to build a multi-frequency model of a mesh with fabrication errors and successfully demonstrated accurate prediction of its behavior. Furthermore, we applied our method on implementations of 6 different kind of FIR/IRR filters, to further prove the effectiveness of our method in configuring applications on meshes with fabrication errors. At last, our method was carried out under various scenarios considering beam splitter splitting ratio variance, inaccurate measurements of mesh and imprecise TBU insertion loss characterization, to demonstrate its strong robustness under various practical scenarios.

## 1. Introduction

Over the past few decades, application-specific photonic integrated circuits (ASPICs) have made significant advancements and shown great potential in various applications [1]. However, up to this point, due to long development period of typically 12 to 24 months per design-fab-packaging-test iteration, only in a few fields like transceivers and data centers, the fabrication volumes are high enough to compensate for the nonrecurring overhead costs [2-4]. As an alternative to ASPICs, programmable photonic integrated circuits (PPICs), inspired by electrical field programmable gate arrays (FPGAs), with general-purpose functionality and flexible operations have emerged as a cost-effective solution [5-15]. At the heart of a PPIC is a general waveguide mesh build from many tunable basic units (TBUs) connected based on two-dimensional (2D) fixed topologies, as shown in Fig. 1(a) and Fig. 1(b). Multiple functionalities can be configured on a same general waveguide mesh by applying voltages on TBUs. Waveguide meshes can be divided into two categories based on whether it support feedback light propagation. The first kind is feedforward-only waveguide meshes, which, as the name suggests, only allow light to propagate forward (a typical structure is shown as in Fig. 1(a)). Such waveguide meshes are often used to configure feedforward multiport interferometers, which are common in quantum optics and artificial intelligence[8, 16]. While this covers a wide range of applications, it does not enable the programming of resonant structure where simultaneous feedforward and feedback light propagation is needed. Thus, the second kind of waveguide meshes, here we call them recirculating waveguide meshes, comes in handy (a typical structure is shown as in Fig. 1(b)). Such waveguide meshes allow for both feedforward and feedback light propagation, thus are the most flexible waveguide meshes can be used to configure both finite (FIR) and infinite (IIR) multiport interferometers and filters[8, 16].

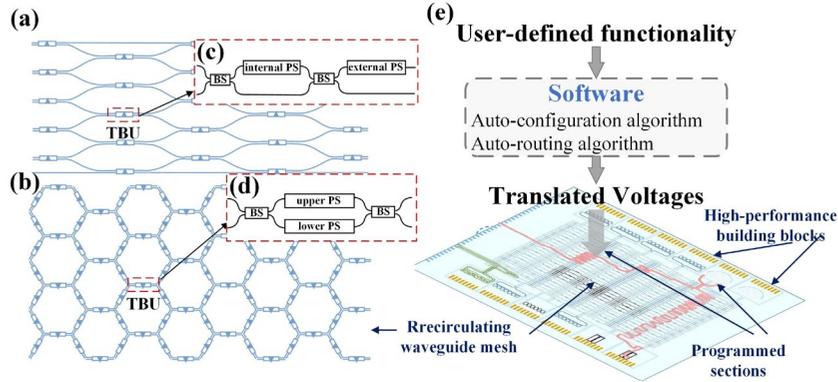

Fig. 1. (a) Feedforward-only waveguide mesh. (b) Recirculating waveguide mesh. (c) Tunable basic unit (TBU) within feedforward-only waveguide mesh, typically implemented with an asymmetric Mach-Zehnder interferometer (MZI), with an internal phase shifter (PS) on one arm and an external PS. (d) TBU within recirculating waveguide mesh, typically implemented with a symmetric MZI with PSs on both arms. (e) Programmable Photonic Circuits (PPICs) based on recirculating waveguide mesh, including a general waveguide mesh, several high-performance building blocks (HPBs), and a software layer. User-defined functionalities are translated by the software layer to voltages applied on TBUs, achieving on-demand light manipulating [14].

In this article, we focus on the field of recirculating waveguide meshes. A complete and practically applicable PPIC should not only include a recirculating waveguide mesh but also several high-performance building blocks (HPBs). Besides the hardware layer, a software layer is essential, as illustrated in Fig. 1(e). This software layer comprises various configuration algorithms responsible for translating user-defined functionalities into voltages applied on TBUs. In this way, light is manipulated as demand to form interference and establish connections to the HPBs, achieving the so-called programming [6, 14]. Current reported configuration algorithms can be divided into two categories. The first one is called auto-configuration algorithm, utilizing optimization algorithms to approximate the working state of the waveguide mesh to meet the user-defined target functionality [14, 17, 18]. The second one is called auto-routing algorithm, utilizing graph theory algorithms to automatically establish connection routes between user-specified ports and construct delay lines [19-22]. However, the imperfections introduced during fabrication may degrade the performance of both categories of algorithms significantly, as the voltages retrieved based on an ideal assumption may not be suitable for the actual fabricated mesh [23, 24]. One possible solution to this problem is to run these algorithms in an on-chip manner, i.e., treating the fabricated mesh as a black-box, obtaining the feedback for the algorithms by measuring the fabricated mesh [10, 14, 17, 25], thus, the actual status of the mesh and its non-idealities can be considered. This ensures setup suitable for the fabricated imperfect mesh, but at the cost of efficiency [17, 18, 26]. For example, without information of the fabricated mesh (treated as a black box), gradient can only be calculated numerically using the measuring results of fabricated mesh, when using gradient-descent optimization to realize configuration. Such numerical gradient calculation involves performing perturbation of voltage on one phase shifter (PS), then measuring the resulting transmission matrix of the fabricated mesh, and repeating the procedure for the next PS, until all PSs in the mesh have been covered. Given the large number of PSs, along with the time-consuming nature of setting and measuring the fabricated mesh, numerical gradient calculation can become time-intensive. Furthermore, the calculation procedure can become even more complex for more accurate gradient calculation [26], letting alone such gradient calculation must be carried out for each optimization iteration, hence, such on-chip manner of configuration is very inefficient.

Therefore, to solve the inefficiency issue, running these two categories of algorithms in an off-chip manner is necessary. So far, simulation methods capable of predicting the mesh transmission matrix [24, 27-29] have been proposed, allowing us to replace the measurements of actual mesh with simulations. Additionally, analytical gradient calculation methods have been proposed, allowing us to replace the time-consuming numerical gradient calculation with a much faster analytical gradient calculation [28, 29]. While these methods represent significant progress towards off-chip configuration, they require being supplied with information of the fabricated mesh to ensure the predicted transmission matrices or gradient are in line with the actual mesh. Therefore, practical off-chip configuration calls for characterization of the fabricated mesh.

In the field of feedforward-only waveguide mesh, many efforts have been made to characterize imperfections, as in [30, 31]. However, these characterization methods cannot be directly applied to recirculating waveguide mesh for the following reasons. First, the structure of TBU in recirculating meshes is different from that in feedforward-only waveguide meshes, and the characterization requirements are higher. In more details, TBUs in feedforward-only waveguide meshes are typically implemented with an asymmetric Mach-Zehnder interferometer (MZI), with an internal phase shifter (PS) on one arm and an external PS [30, 31], as shown in Fig. 1(c). In contrast, TBUs in recirculating waveguide meshes, which function as building blocks for constructing resonators, require a more compact structure with independent control of coupling ratio and phase delay, to provide greater flexibility. Thus, they are typically implemented with a symmetric MZI with PSs on both arms to allow dual-driven [8, 32], as shown in Fig. 1(d). This requires the actual phases of the PSs on both the upper and lower arms of MZI to be characterized, whereas in feedforward-only waveguide meshes, the primary concern is the phase difference between the upper arm PS and the lower arm of MZI. The phase difference can be easily deduced from the interference output light intensity [30, 31]. But the actual phase of each PS cannot be deduced. Second, the more complex connection topology of recirculating waveguide meshes further complicated its characterization. In feedforward-only waveguide meshes, the phase difference can be easily obtained by forming interference and deducing from the detected interference light intensity [30, 31]. In recirculating waveguide meshes, however, the formation of interference typically requires the participation of multiple TBUs duo to their complex topology. This makes it difficult to analytically separate the effects of different TBU, thus making it impossible to deduce phase information. Finally, compared to feedforward-only waveguide meshes, the performance of recirculating waveguide meshes needs to be evaluated at more frequency points since they are often used to configure filters. In contrast, feedforward-only meshes typically operate at a single frequency point because they are commonly used to build multiport interferometers in quantum optics and artificial intelligence. Consequently, the characterization of recirculating waveguide meshes is particularly challenging. As a result, characterization methods for recirculating waveguide meshes proposed so far primarily focus on characterizing only the coupling factor of TBU relative to applied voltage[25, 26, 33], which is just one of the TBU performance indices. Other indices, such as passive phase and phase increment relative to voltage of its PS, splitting ratios of its BS, group index, etc., are also crucial for the operation of the simulation method and analytical gradient calculation method.

Therefore, in this paper, we propose a method capable of characterize all these performance indices of each TBU in the recirculating waveguide mesh. This is done by employing an optimization method to find the most suitable values for these imperfect parameters, based on the criterion of making the mesh behavior closely resembles that of the fabricated one. As explained before, the lack of information of phase could making the characterization of recirculating waveguide meshes particularly challenging, thus we also develop a step-by-step procedure to disentangle various imperfect parameters and narrow down their potential ranges before optimization, especially in step 3, we propose a novel method to narrow down the value range of phase of PS to just two potential values (Step 3), this helps to narrow down the parameter space and provide a good starting position for optimization. Also, since recirculating waveguide meshes often used to configure filters, we include characterizing the group index of waveguide in our characterization method, to allow evaluating the performance of mesh over the entire frequency band, rather than just a single frequency point. To verify the effectiveness of our proposed method, we used the characterized parameters to build a model of a mesh with fabrication errors and demonstrated its ability in accurately predicting the mesh transmission matrices. Moreover, we carried out our method under various scenarios considering beam splitter splitting ratio variance, inaccurate measurements of mesh and imprecise TBU insertion loss characterization, to analyze the stability and robustness of our method. At last, we applied our method on implementations of 6 different kind of finite impulse response (FIR) filters or infinite impulse response (IIR) filters, to further prove the effectiveness of our method in configuring applications on meshes with fabrication errors. Our method provides the last crucial element needed to achieve off-chip configuration. It can free the configuration process from the time-consuming task of setting and measuring the fabricated mesh. Instead, it enables the entire configuration process to be conducted solely through data calculations.

## 2. Modeling of recirculating waveguide mesh considering fabrication imperfection

Recirculating waveguide meshes rely on the interconnection of TBUs (as shown in Fig. 2(a)), which is complemented with a symmetric MZI composed of 50:50 beam splitters (BSs) and two phase shifters (PSs) attached to both arms, as shown in Fig. 2(b).

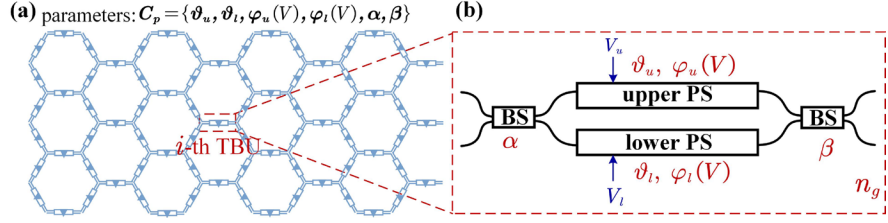

Fig. 2. (a) Recirculating waveguide mesh consisting of interconnected TBUs (b) TBU employing Mach-Zehnder interferometer (MZI) structure, composed of beam splitters (BSs) and phase shifters (PSs). $\vartheta_u$ and $\vartheta_l$ are the actual passive phase of the PS on the upper and lower arm of MZI. $\varphi_u(V)$ and $\varphi_l(V)$ are the actual phase-voltage relationship curves (P-V curves) of the upper and lower PS. $\alpha$ and $\beta$ indicate the actual splitting ratio of BSs. $n_g$ indicated the actual group index of waveguide. $V_u$ and $V_l$ are the voltages applied on the upper and lower PS.

Due to fabrication imperfections, the actual values of these parameters may deviate from the ideal one. Thus, the scattering matrix of an actual imperfect TBU can be expressed as:

$$S_{TBU} = \begin{bmatrix} \cos(\beta) & j\sin(\beta) \\ j\sin(\beta) & \cos(\beta) \end{bmatrix} \begin{bmatrix} e^{j\theta_u} & 0 \\ 0 & e^{j\theta_l} \end{bmatrix} \begin{bmatrix} \cos(\alpha) & j\sin(\alpha) \\ j\sin(\alpha) & \cos(\alpha) \end{bmatrix} \quad (1)$$

where, $\alpha$ and $\beta$ indicate the actual splitting ratio of BSs. $\theta_u/\theta_l$ is the actual phase shift of the upper/lower PS when apply voltage $V_u/V_l$ on it. Considering recirculating waveguide meshes are often used to configuring filters, it is not enough to simply evaluating its performance at a single frequency point. Thus, we perform a first-order Taylor expansion of $\theta_u$ and $\theta_l$ with respect to optical frequency $f$, providing a model that is applicable across a broader frequency band. $\theta_u$ and $\theta_l$ can be expressed as:

$$\theta_u(f, V_u) = \vartheta_u + \frac{2\pi}{c} L n_g (f - f_0) + \varphi_u(V_u) \quad (2)$$

$$\theta_l(f, V_l) = \vartheta_l + \frac{2\pi}{c} L n_g (f - f_0) + \varphi_l(V_l) \quad (3)$$

where, $f_0$ is the reference frequency, $n_g$ is the actual group index of waveguide, $L$ is the length of the PS. $\vartheta_u$ and $\vartheta_l$ are the actual passive phase (without voltage applied) of the upper and lower PS at reference frequency $f_0$, $\varphi_u(V)$ and $\varphi_l(V)$ are the actual phase-voltage relationship curves (P-V curves) of the upper and lower PS, as labeled in Fig. 2(b). From Eq. (1), Eq. (2) and Eq. (3), it can be observed that when parameters $\vartheta_u, \vartheta_l, \varphi_u(V), \varphi_l(V), n_g, \alpha$, and $\beta$ are known, the scattering matrix of TBU can be deduced. Therefore, once we characterize $\boldsymbol{C_p} = \{\boldsymbol{\vartheta_u}, \boldsymbol{\vartheta_l}, \boldsymbol{\varphi_u}(V), \boldsymbol{\varphi_l}(V), \boldsymbol{\alpha}, \boldsymbol{\beta}\}$, i.e., the actual value of these parameters of each TBU in the fabricated mesh, the scattering matrix of the waveguide mesh can then be deduced using methods proposed in [24, 27-29], thus the behavior of the actual fabricated waveguide mesh under any voltage setting $\boldsymbol{V} = [V_{u,1}, V_{l,1}, V_{u,2}, V_{l,2}, \cdots, V_{u,N_{TBU}}, V_{l,N_{TBU}}]$, where $N_{TBU}$ indicates the number of TBUs in the mesh, can be predicted. $\boldsymbol{\vartheta_u} = [\vartheta_{u,1}, \vartheta_{u,2}, \cdots, \vartheta_{u,N_{TBU}}]$, $\boldsymbol{\vartheta_l} = [\vartheta_{l,1}, \vartheta_{l,2}, \cdots, \vartheta_{l,N_{TBU}}]$, $\boldsymbol{\varphi_u}(V) = [\varphi_{u,1}(V), \varphi_{u,2}(V), \cdots, \varphi_{u,N_{TBU}}(V)]$, $\boldsymbol{\varphi_l}(V) = [\varphi_{l,1}(V), \varphi_{l,2}(V), \cdots, \varphi_{l,N_{TBU}}(V)]$, $\boldsymbol{\alpha} = [\alpha_1, \alpha_2, \cdots, \alpha_{N_{TBU}}]$, $\boldsymbol{\beta} = [\beta_1, \beta_2, \cdots, \beta_{N_{TBU}}]$, and $\boldsymbol{n_g} = [n_{g,1}, n_{g,2}, \cdots, n_{g,N_{TBU}}]$ are vectors composed of their corresponding parameters of each TBU in the mesh. Subscripts are used to indicate the parameter upper/lower location within TBU, and the corresponding TBU identifier. Bold letters indicate vectors, lowercase letters are used for their elements.

## 3. Principle of proposed characterization method

Due to complex topology of recirculating waveguide meshes, components with imperfections are highly coupled, making it challenging to analytically disentangle and characterize all parameters in $\boldsymbol{C_p}$. Therefore, we adopt an optimization method to characterize them, i.e., using an optimization method to find the most suitable value for $C_p$, based on the criterion of making the mesh behavior closely resembles that of the fabricated one. To do so, we define a cost function CF, as in Eq. (4),

$$CF = \sum_{l=1}^{N_V} \sum_{k=1}^{N_f} \sum_{j=1}^{N_{port}} \sum_{i=1}^{N_{port}} \left( \sqrt{\hat{t}_{i,j,k}^{(l)}} - \sqrt{t_{i,j,k}^{(l)}} \right)^2 \tag{4}$$

which measures the difference between the transmission matrices of the actual fabricated mesh, denoted as $\boldsymbol{T}^{(l)}$, and the transmission matrices would manifest under current assumed parameter $\hat{\boldsymbol{C}}_p$, denoted as $\hat{\boldsymbol{T}}^{(l)}$, under various voltage settings. $t_{i,j,k}^{(l)}$ and $\hat{t}_{i,j,k}^{(l)}$ are the weights of $\boldsymbol{T}^{(l)}$ and $\hat{\boldsymbol{T}}^{(l)}$, respectively. The subscripts $i$, $j$ and $k$ represent the row, column, and spectral slice index, respectively. The superscript $l$ denote the index for a particular voltage setting. $N_{port}$ is the number of ports of the waveguide mesh, determining the row and column count of $\boldsymbol{T}^{(l)}$, $N_f$ refers to the spectral slice count of $\boldsymbol{T}^{(l)}$, which is the number of frequency points at which the transmission matric is measured after each voltage application. $N_V$ is the total number of voltage settings.

The optimal value of $C_p$ will be found when CF is minimized, and $C_p$ is thus characterized. Actually, instead of directly optimizing all parameters in $C_p$, we develop a step-by-step procedure to isolate and determine the values of specific parameters before optimization, particularly in Step 3, we propose a novel method capable of narrowing down the value range of passive phase of PS to just two potential values, thereby helping to simplify the optimization by reducing the parameters space beforehand.

The characterization procedure can be divided into 4 steps.

### Step 1: Characterization of $d\vartheta = \vartheta_l - \vartheta_u$, $\varphi_u(V)$ and $\varphi_l(V)$

We can have $d\vartheta = \vartheta_l - \vartheta_u$, $\varphi_u(V)$ and $\varphi_l(V)$ through the coupling factor of TBU relative to the applied voltages ($V$), which can be obtained based on the method proposed in [33].

### Step 2: Characterization of $n_g$

To characterize $n_g$, we synthesize a MZI resonator on the mesh by setting certain TBUs to cross, bar, or 50:50 coupling state (the voltage values needed for TBUs to reach their target coupling state is known after Step 1). Then we extract Free Spectral Range (FSR) from the resonance spectrum and calculate $n_g$ using Eq. (5).

$$n_g = \frac{c}{10 \cdot L \cdot FSR} \tag{5}$$

Note that, we assume a same value of $n_g$ for all TBUs, since the difference of group refractive index among waveguides within the same region is not significant [34, 35].

### Step 3: Narrowing down value range of $\vartheta_u$ and $\vartheta_l$

As explained before, in recirculating waveguide meshes, gaining information about the phase of PS is difficult, making the characterization of recirculating waveguide meshes particularly challenging. Here, we propose a novel method capable of narrowing down the value range of passive phase of PS ($\theta_{u/l}$) to two potential values, helping to reduce the parameter space prior to optimization. This method ingeniously constructs two delay lines, allowing for the extraction of sum of $\vartheta_u$ and $\vartheta_l$ from the phase difference between the two delay lines. Combine with the information of $d\vartheta = \vartheta_l - \vartheta_u$, which we have already obtained in Step 1, the value of $\vartheta_u$ and $\vartheta_l$ can then be solved. The detailed principle of the method is provided as follows.

Assuming perfect splitting ratio $K_{BS} = 50\%$ for BSs, thus Eq. (1) can be written as:

$$S_{TBU} = e^{j\phi} \begin{bmatrix} \sin\Delta & \cos\Delta \\ \cos\Delta & -\sin\Delta \end{bmatrix} \tag{6}$$

in which,

$$\Delta = \frac{\theta_u - \theta_l}{2} = \frac{\vartheta_u - \vartheta_l + \varphi_u(V_u) - \varphi_l(V_l)}{2} \quad (7)$$

$$\phi = \frac{\pi}{2} + \frac{\theta_u + \theta_l}{2} = \frac{\pi}{2} + \frac{\vartheta_u + \vartheta_l + \varphi_u(V_u) + \varphi_l(V_l)}{2} \quad (8)$$

As we can see, $\phi$ contains the information of $\vartheta_u + \vartheta_l$. If we know $\phi$, plus that we have already characterize $d\vartheta = \vartheta_l - \vartheta_u$, we can solve for the specific values of $\vartheta_u$ and $\vartheta_l$.

We use $\phi^{cross}$, corresponding to $\phi$ when TBU is set to cross state, to calculate $\vartheta_u + \vartheta_l$. When keeping the voltage on the lower PS to be zero, and apply voltage $V_u^{cross}$ on the upper PS to set TBU to cross state, Eq. (7) and Eq. (8) can be written as:

$$\Delta^{cross} = \frac{\vartheta_u - \vartheta_l + \varphi_u(V_u^{cross})}{2} \quad (9)$$

$$\phi^{cross} = \frac{\pi}{2} + \frac{\vartheta_u + \vartheta_l + \varphi_u(V_u^{cross})}{2} \quad (10)$$

As a matter of fact, obtaining the actual value of $\phi^{cross}$ is challenging, but the modulus of $2\phi^{cross}$ in radians can be easily obtained. Thus, we reorganize Eq. (10) as:

$$\vartheta_u + \vartheta_l = mod(2\phi^{corss}) - \pi + 2m\pi - \varphi_u(V_u^{cross}) \quad (11)$$

$m \in \mathbf{Z}$. $mod$ means taking the modulus in radians. When TBU is at cross state, $\Delta^{cross}$ should satisfy $\Delta^{cross} = n\pi$, substitute into Eq. (9), we get:

$$\vartheta_u - \vartheta_l + \varphi_u(V_u^{cross}) = 2n\pi \quad (12)$$

$n \in \mathbf{Z}$. Combining Eq. (11) and Eq. (12), as well as $d\vartheta = \vartheta_l - \vartheta_u$, we get:

$$\vartheta_u + \vartheta_l = mod(2\phi^{cross}) - d\vartheta - \pi + 2(m-n)\pi \quad (13)$$

$$\vartheta_u - \vartheta_l = -d\vartheta \quad (14)$$

Thus $\vartheta_u$ and $\vartheta_l$ can be solved as:

$$\vartheta_u = \frac{1}{2}mod(2\phi^{cross}) - d\vartheta - \frac{\pi}{2} + (m-n)\pi \quad (15)$$

$$\vartheta_l = \vartheta_u + d\vartheta \quad (16)$$

where m and n are chosen so that $\vartheta_u$ are in the range of $[0, 2\pi)$. In fact, $\vartheta_u$ has two solutions, thus, we have narrowed down the possible range of $\vartheta_u$ to two possibilities.

From Eq. (15) and Eq. (16), we can see that, $\vartheta_u$ and $\vartheta_l$ can be solved once we obtain $mod(2\phi^{cross})$. To obtain the value of $mod(2\phi^{cross})$, we designed a MZI resonator, the upper arm of which is an external phase shifter, and the lower arm is a delay line configured on the waveguide mesh, as shown in Fig. 3(a). With different settings of the waveguide mesh, the delay line light path shapes either "8" or "0", as shown in Fig. 3(b) and Fig. 3(c).

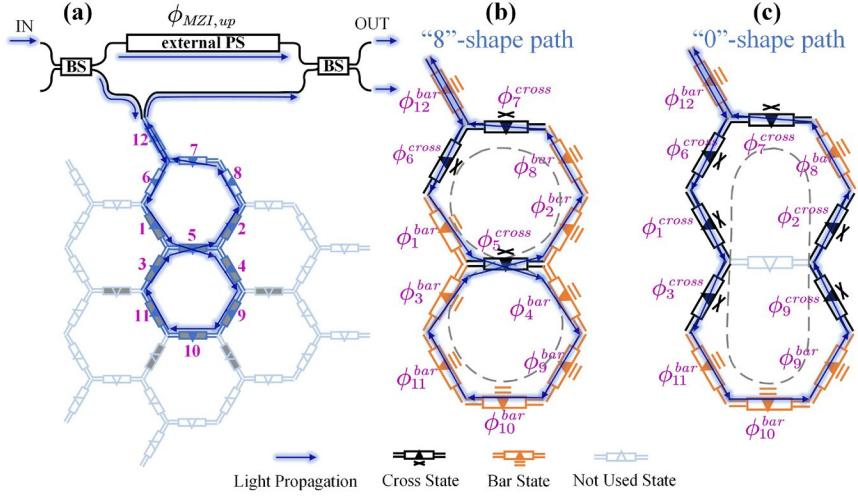

Fig. 3. (a) Constructed MZI resonator, the upper arm of which is an external phase shifter, and the lower arm is a delay line configured on the waveguide mesh. The TBU identifiers are marked with purple numbers. A delay line configured on the waveguide mesh, shaping like (b) "8", and (c) "0". $\phi_i^{st}$ is the corresponding $\phi$ of the i-th TBU in $st$ state ($st \in \{cross, bar\}$); $\phi_{MZI,up}$ represents the phase of the upper arm of the constructed MZI.

$\phi_{"8"}$ and $\phi_{"0"}$, represent the total phase shift of the "8"-shape and "0"-shape delay lines, respectively, can be expressed as sum of $\phi$ of their constituted TBUs:

$$\phi_{"8"} = \sum_{i=1}^{4} \phi_i^{bar} + \sum_{i=6}^{7} \phi_i^{cross} + \sum_{i=8}^{12} \phi_i^{bar} + 2\phi_5^{cross} \tag{17}$$

$$\phi_{"0"} = \sum_{i=1}^{4} \phi_i^{cross} + \sum_{i=6}^{7} \phi_i^{cross} + \sum_{i=8}^{12} \phi_i^{bar} \tag{18}$$

where $\phi_i^{st}$ is the corresponding $\phi$ of the i-th TBU in $st$ state ($st \in \{cross, bar\}$). From Eq. (9) and Eq. (10) we can deduce that $\phi_i^{bar} - \phi_i^{cross} = \pi/2$.

Subtract Eq. (17) and Eq. (18), and substitute $\phi_i^{bar} - \phi_i^{cross} = \pi/2$, we get:

$$mod(2\phi_5^{cross}) = mod(\phi_{"8"} - \phi_{"0"}) \tag{19}$$

It can be observed in Eq. (19) that we can have $mod(2\phi_5^{cross})$ once we know $\phi_{"8"}$ and $\phi_{"0"}$. However, obtaining the absolute values of $\phi_{"8"}$ and $\phi_{"0"}$ is still difficult. But we can have $mod(\phi_{"8"} - \phi_{"0"})$ by acquiring its equal value $mod(\delta\phi_{"8"} - \delta\phi_{"0"})$, $\delta\phi_{"8"}$ and $\delta\phi_{"0"}$ are the phase differences between the two arms of the constructed MZI, respectively corresponding to using the "8"-shape and the "0"-shape delay line as its lower arm. $\phi_{"8"}$ and $\phi_{"0"}$ are related to $\delta\phi_{"8"}$ and $\delta\phi_{"0"}$ as in Eq. (20) and Eq. (21).

$$\delta\phi_{"8"} = mod(\phi_{"8"} - \phi_{MZI,up}) \tag{20}$$

$$\delta\phi_{"0"} = mod(\phi_{"0"} - \phi_{MZI,up}) \tag{21}$$

$\phi_{MZI,up}$ represents the phase of the upper arm of the constructed MZI

Subtracting Eq. (20) and Eq. (21), $mod(\phi_{"8"} - \phi_{"0"})$ can be expressed as:

$$mod(\phi_{"8"} - \phi_{"0"}) = mod(\delta\phi_{"8"} - \delta\phi_{"0"}) \tag{22}$$

$\delta\phi_{"8"}$ and $\delta\phi_{"0"}$ can be easily obtained. Sweeping the voltage applied on the external PS of the constructed MZI, while testing the output power of its output port, $\delta\phi_{"8"}$ and $\delta\phi_{"0"}$ can be easily extracted from the power-voltage curve. Thus, according to Eq. (22) and Eq. (19), we now have $mod(2\phi^{cross})$, $\vartheta_u$ and of $\vartheta_l$ TBU 5 can then be solved using Eq. (15) and Eq. (16). Similarly, for all

TBUs that are not on the edge, painted gray as in Fig. 3(a), we can perform the above operation to solve their $\vartheta_u$ and $\vartheta_l$.

*Step 4: characterize $C_p$ utilizing optimization*

So far, the remaining parameters in $C_p$ that have not been characterized are $\vartheta_u, \vartheta_l, \alpha$ and $\beta$. Actually we only need to take $\vartheta_u$, $\alpha$ and $\beta$ as variables, since $\vartheta_l$ can be calculated using $\vartheta_l = \vartheta_u + d\vartheta$, and $d\vartheta$ has already been characterized in Step 1. Here we also include $d\vartheta$ in optimization to characterize it further accurate, since the value obtained after Step 1 may still contain errors. So, the final variables we choose to include in optimization are $C_p' = \{\vartheta_u, d\vartheta, \alpha, \beta\}$.

The workflow of the optimization is illustrated in Fig. 4(a).

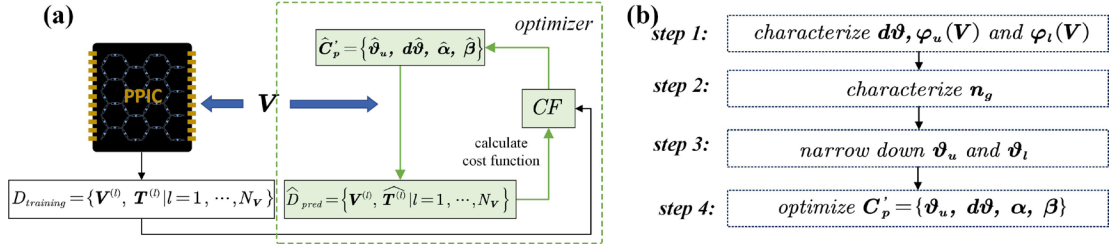

Fig. 4. (a) Optimization system diagram. (b) Characterization procedure involving 4 steps. Step 1: characterize passive phase difference $d\vartheta = \vartheta_l - \vartheta_u$ between the upper and lower PS, and the phase-voltage relationship curves $\varphi_u(V)/\varphi_l(V)$ of the upper and lower PS. Step 2: characterize the group index $n_g$ of waveguide. Step 3: narrow down the potential values of $\vartheta_u$ and $\vartheta_l$. Step 4: optimize $C_p' = \{\vartheta_u, d\vartheta, \alpha, \beta\}$.

Randomly generate $N_V$ sets of $V$, measure the transmission matrix of the actual fabricated mesh $T^{(l)}$ under each voltage setting $V^{(l)}$, forming a dataset $D_{training} = \{V^{(l)}, T^{(l)} | l = 1, \cdots, N_V\}$. Such dataset describes the behavior of the fabricated mesh under various voltage settings. For each voltage setting, we calculate the corresponding transmission matrix $\hat{T}^{(l)}$ would manifest under current assuming $\hat{C}_p'$, forming a dataset $\hat{D}_{pred} = \{V^{(l)}, \hat{T}^{(l)} | l = 1, \cdots, N_V\}$. Such dataset describes the behavior of a mesh with current assuming $\hat{C}_p'$. Then we calculate the CF, using Eq.(4), to measure the difference between $D_{training}$ and $\hat{D}_{pred}$, in other words, to determine whether a mesh with current assuming $\hat{C}_p'$ would behave similarly to the actual mesh. Iterating to find the optimal value of $\hat{C}_p'$, which will be reached when CF is minimized.

Given that we are dealing with an optimization problem characterized by continuous space, high dimensionality, non-convexity, and a solution space featuring numerous hills and valleys, we choose to employ Particle Swarm Optimization (PSO) algorithm [36, 37] to conduct optimization, which is well-suited for addressing such challenges.

After completing all four steps, $\hat{C}_p = \{\hat{\vartheta}_u, \hat{\vartheta}_l, \hat{\varphi}_u(V), \hat{\varphi}_l(V), \hat{\alpha}, \hat{\beta}\}$ can be synthesized from their results. To summarize, the characterization procedure involving the 4 steps is illustrated in Fig. 4(b).

## 4. Simulation Results

Firstly, in order to verify the effectiveness of our method in accurately characterizing the imperfections, we compared the characterized values with the actual ones. We also used the characterized parameters to build a multi-frequency model of the imperfect mesh, demonstrating its ability in accurately predicting mesh transmission matrices. Secondly, we conducted our method under various practical scenarios to evaluate its robustness. At last, we applied our method on implementations of 6 FIR/IRR filters, to further demonstrate its effectiveness in configuring applications on mesh with fabrication errors.

*4.1 Characterization errors and accuracy of predicting mesh transmission matrices*

We conducted simulation verification of our method on a waveguide mesh containing 36 TBUs. Considering imperfections, splitting ratios of BSs, denoted using $K_{BS}$, follow a Gaussian distribution with a mean of 50% and a standard deviation of $\sigma_{BS} = 2.5\%$ (typical wafer-level variation of $\sigma_{BS} = 2\%$ [38]). $\vartheta_u$ and $\vartheta_l$ follow uniform distributions within the range of $[0, 2\pi)$, $\varphi_u(V)$ and $\varphi_l(V)$ are proportional to square of applied voltage $V$. $n_g$ follows a Gaussian distribution with a mean of 4.3 and a standard deviation of 0.01 [34, 35]. Based on this, we randomly generate imperfect parameter values ($\boldsymbol{C_p}$), and then employ the proposed characterization method to try to retrieve them. When conducting the optimization in Step 4, we choose $N_{\boldsymbol{V}} = 45$ and $N_f = 1$ to generate $D_{training}$.

Here, we show the characterization results. After Step 3, the value range of $\vartheta_u$ of a subset of TBUs have been narrowed down to two potential values. As shown in Fig. 5(a), the actual value of $\vartheta_u$ are indeed very close to one of the predicted potential values. The subtle difference between them arises from that, the calculation of these two potential values uses the assumption of $K_{BS} = 50\%$, which cannot be guaranteed due to fabrication errors.

We conducted 10 experiments by randomly generating 10 sets of different $\boldsymbol{C_p}$ values, then use our characterization method to respectively retrieve them, and calculate the characterization errors. Fig. 5(b) shows the probability density function (PDF) of characterization error of $K_{BS}$, indicating that our method ensures characterizing $K_{BS}$ of BSs with an error less than 1.34% in 95% of the case. Fig. 5(c) shows the PDF of characterization error of $d\vartheta$, indicating that our method can ensure characterizing $d\vartheta$ with an error smaller than $0.0026\pi$ in 95% of the case. To evaluate the effectiveness of our method in ensuring accurate modeling of the actual mesh, we use the characterized parameters to build a model of the actual mesh, and evaluate its accuracy in predicting the mesh transmission matrices. In order to demonstrate performance of such model across multiple frequency points and varying voltage settings, we use such model to predict transmission matrices (in dB) at $N_f = 50$ frequency points across the whole FSR (corresponding to the length of a single TBU), respectively under $N_{\boldsymbol{V}} = 100$ sets of $\boldsymbol{V}$ settings, and compare the predicted transmission matrices ($\hat{\boldsymbol{T}}$) with those of the actual mesh ($\boldsymbol{T}$). The PDF of the prediction error is plot in Fig. 5(d), indicating that our method can ensure accurate prediction of transmission matrices with an error less than 0.55dB for 85% of the case.

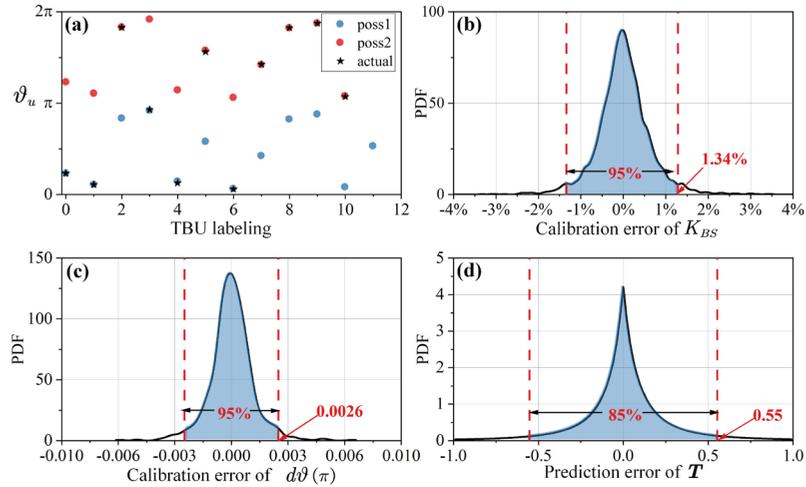

Fig. 5. (a) Compare the actual value of $\vartheta_u$ (labeled as "actual") with the two potential values calculated in Step 3 (labeled as "poss1" and "poss2"). Probability density functions (PDFs) of the characterization error of (b) $K_{BS}$ and (c) $d\vartheta$. (d) PDF of prediction error of transmission matrix $\boldsymbol{T}$.

We also adopt the root-mean-square error (RMSE) (Eq. (23)) to measure the average prediction error of the transmission matrices. In our case, the average RMSE over 10 experiments is 0.34 dB

$$RMSE = \sqrt{\frac{\sum_{l=1}^{N_V}\sum_{k=1}^{N_f}\sum_{j=1}^{N_{port}}\sum_{i=1}^{N_{port}}\left(10\lg\left(\widehat{t_{i,j,k}^{(l)}}\right) - 10\lg\left(t_{i,j,k}^{(l)}\right)\right)^2}{N_{port}^2 \cdot N_V \cdot N_f}} \qquad (23)$$

### 4.2 Stability and Robustness

In this section, we consider various practical scenarios, evaluating the performance of our proposed characterization method under different conditions.

We first evaluate the stability and robustness of our characterization method under different $K_{BS}$ variation scenario. Considering standard deviation of $K_{BS}$, denoted as $\sigma_{BS}$ to be 2.5%, 5%, 7.5%, respectively. Under each scenario, conduct 10 experiments by randomly generating 10 sets of different $C_p$ values, and use our characterization method to respectively retrieve them. The resulting RMSEs are depicted in Fig. 6(a). Notably, our characterization method proves effective even under extreme splitting ratios, which is way worse than the typical wafer-level variance (beam splitter variation as small as 2% [38]), and is very stable, cross all experimental runs for various splitting ratio conditions, the RMSEs stay below 1.0 dB.

Next, we considering measurement inaccuracy when measuring the transmission matrices $T$ of the fabricated mesh, we simulate the measurement inaccuracy by introducing random fluctuations to the accurate $T$, where the magnitude of fluctuations follows Gaussian distributions with standard deviations $\sigma_T$ of 1%, 2%, and 3%, repectively. Under each scenario, conducted 10 experiments. The resulting RMSEs are depicted in Fig. 6(b). The performance of the characterization method indeed decreases with the rise in measurement inaccuracy, but the decrease is small. Our method maintains good performance even under vary bad conditions.

In the previously validation process, we did not consider the insertion loss (IL) of TBUs, as IL of each TBU can be characterized using the method outlined in [26]. When characterized accurately, ILs of the TBUs would not impact the effectiveness of our method. However, acknowledging potential inaccuracies in IL characterization, as described in [26], where the characterized ILs have an average error of 0.18 dB, here we evaluate the robustness of our characterization method against IL characterization errors. Assuming the actual ILs of TBUs are uniformly distributed within the range of (0.5, 0.7) dB [39]. Considering the IL characterization errors uniformly distributed within the ranges of (-0.2,0.2) dB, (-0.4,0.4) dB, and (-0.6, 0.6) dB, corresponding to average absolute error of 0.1dB, 0.2dB and 0.3dB, respectively. Under each scenario, conducted 10 experiments. The resulting RMSEs are depicted in Fig. 6(c).

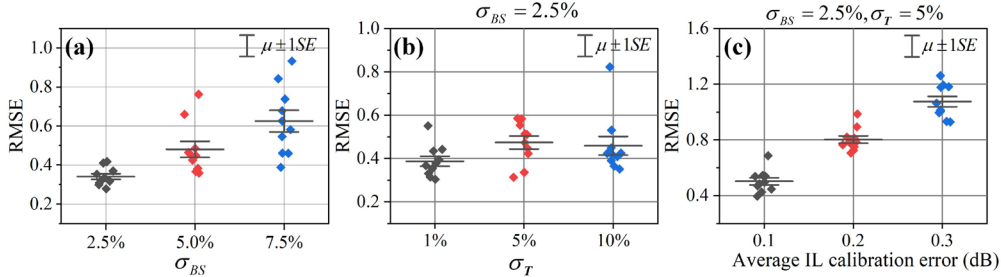

Fig. 6. (a) RMSEs obtained under various $\sigma_{BS}$ conditions. (b) RMSEs obtained under different levels of measurement inaccuracy of $T$, under the condition of $\sigma_{BS}=2.5\%$. (c) RMSEs obtained under different IL characterization error levels, under the condition of $\sigma_{BS}=2.5\%$, $\sigma_T=5\%$. In each scenario, conduct 10 experiments. Plot all obtained RMSE values, and mark the Mean plus Standard Deviation ($\mu+1SE$, $\mu$ indicates mean value, $SE$ indicates standard deviation).

### 4.3 Verification of the proposed method by implementing various applications

At last, we demonstrate the effectiveness of our characterization method through the implementations of finite response (FIR) filter applications, including two MZIs with different arm lengths difference and a 3-tap MZI lattice filter, as shown in Fig. 7(a)(b)(c).

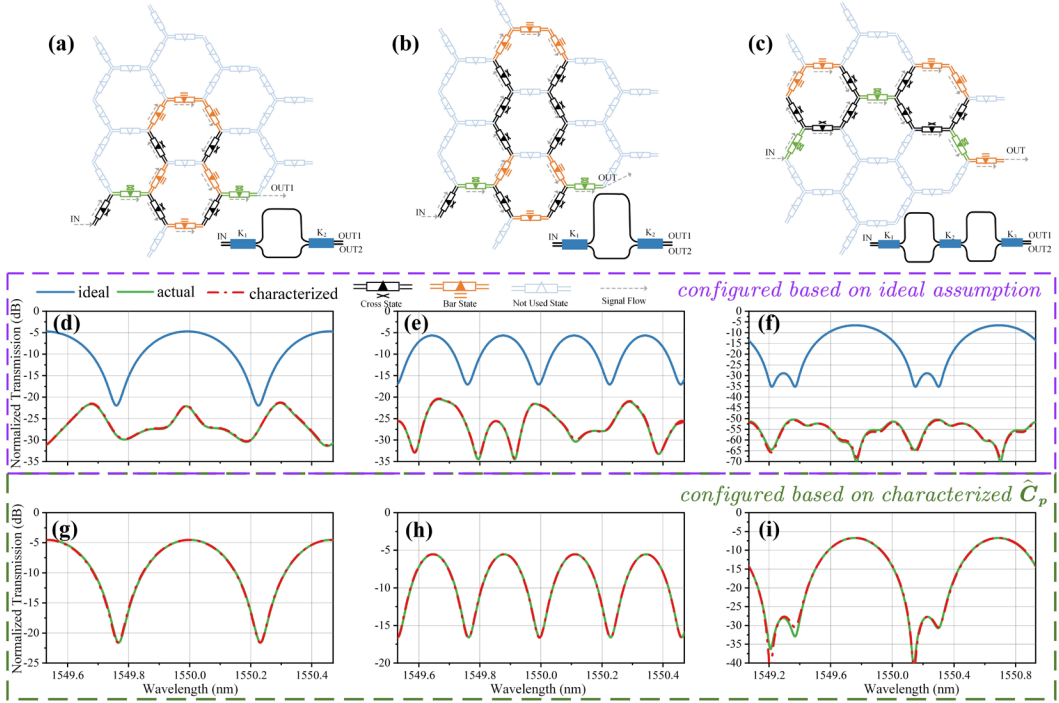

Fig. 7. FIR applications: Circuit layout diagram, waveguide mesh arrangements for three different FIR applications. (a)(b) Mach-Zhender interferometers (UMZIs) with different arm lengths difference. (c) a 3-tap MZI lattice filter. (d)(e)(f) spectral responses of the ideal, actual, and characterized mesh, when configured with voltages chosen based on ideal assumption, respectively, for the three applications. (g)(h)(i) spectral response of the actual and characterized mesh, when configured with voltages chosen base on the characterized $\hat{C}_p$, respectively, for the three applications.

First, we configured the filters using voltages chosen based on an ideal assumption, as a result, the normalized spectral responses of an ideal mesh, the actual mesh, and the characterized mesh were plotted in Fig. 7(d)(e)(f). As we can see, the spectral responses of the actual mesh deviate significantly from those of the ideal mesh, indicating that, configuring the mesh based on an ideal assumption would lead the actual mesh to deviate from the targeted functionality. Also, the spectral responses of the characterized mesh align with those of the actual mesh, validating the effectiveness of our characterization method in accurately predicting the actual mesh behavior. Then we configure the filters using voltages chosen based on the characterized $\hat{C}_p$, the normalized spectral responses of the actual and the characterized mesh are plotted in Fig. 7(g)(h)(i), as we can see, they are in perfect alignment, and both have achieved the intended functionality, highlighting that our characterization method can ensure configurations suitable for the actual mesh. Note that $\hat{C}_p$ used here is characterized under the circumstance of $\sigma_{BS}=2.5\%$, $\sigma_T=5\%$, exhibiting an RMSE of 0.58 dB.

We also demonstrated the implementations of infinite impulse response (IIR) filter applications, including an optical ring resonator (ORR), a triple ORR coupled resonator waveguide (CROW) filter and a double ORR ring-loaded MZI. Similarly, we respectively configured these IIR filters using voltages chosen based on an ideal assumption (corresponding spectral responses plotted in Fig. 8(d)(e)(f)) and voltages chosen based on the characterized $\hat{C}_p$ (corresponding spectral responses plotted in Fig. 8(g)(h)(i)), as shown in Fig. 8, we can observe a similar phenomenon as in the FIR filters configurations. The successfully implementations of the FIR and IIR filters prove the effectiveness of our method in configuring applications on meshes with fabrication errors.

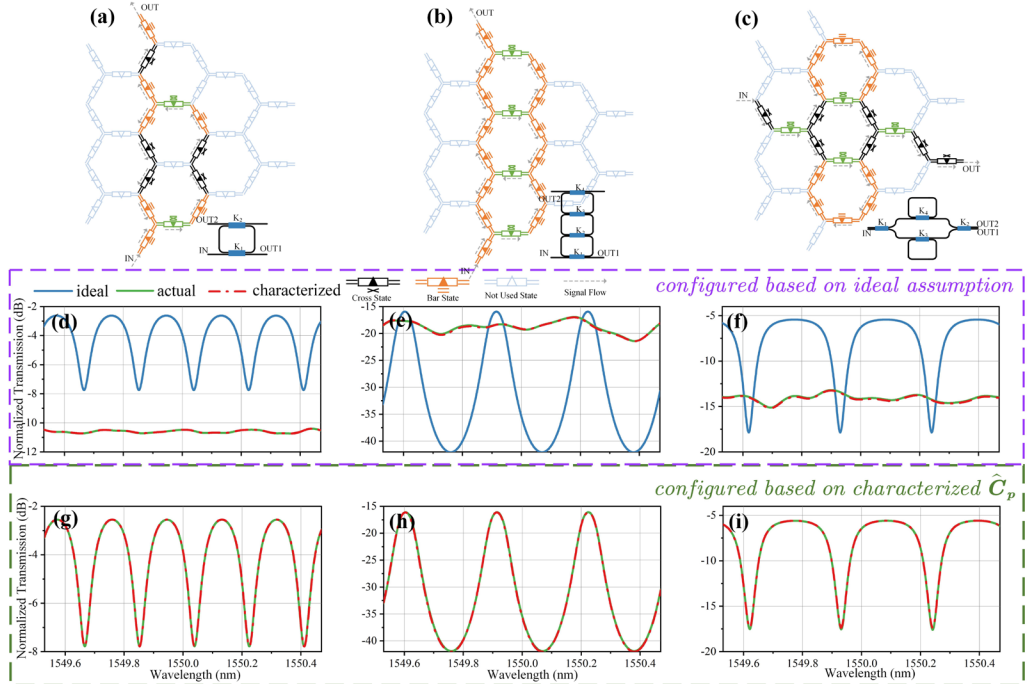

Fig. 8. IIR applications: Circuit layout diagram, waveguide mesh arrangements for three different IIR applications. (a) an optical ring resonator (ORR). (b) a triple ORR coupled resonator waveguide (CROW) filter. (c) a double ORR ring-loaded MZI. (d)(e)(f) spectral responses of the ideal, actual, and characterized mesh, when configured with voltages chosen based on ideal assumption, respectively, for the three applications. (g)(h)(i) spectral response of the actual and characterized mesh, when configured with voltages chosen base on characterized $\widehat{C}_p$, respectively, for the three applications.

## 5. Conclusion

We propose a characterization method for recirculating waveguide meshes based on an optimization approach, along with a procedure to reduce the parameter space prior to optimization, allowing for characterizing imperfect parameters of each individual component of the mesh. Under the condition of $\sigma_{BS}=2.5\%$, our method ensures characterizing $K_{BS}$ of BSs with an error less than 1.34% in 95% of the case, characterizing $d\vartheta_u$ with an error smaller than $0.0026\pi$ in 95% of the case. We also demonstrate that our method ensures accurate prediction of the behavior of mesh with fabrication errors, simulation model build with characterized parameters can ensure predicting the mesh transmission matrices with an average RMSE of 0.34dB. It can be seen that even though our characterization procedure was conducted at a single frequency point, it still allows for the construction of an accurate multi-frequency model of the mesh. We also analyzed the stability and robustness of the proposed characterization method under various scenarios considering $K_{BS}$ variance, inaccuracies in measurements of mesh, and imprecise TBU IL characterization. It has been demonstrated that our characterization method is very robust, exhibits an average RMSE of 0.8dB, under the circumstance of $\sigma_{BS}=2.5\%$, $\sigma_T=5\%$, and average IL characterization error of 0.2dB. We also apply our method on implementations of various FIR and IIR applications, confirming the effectiveness of our method in configuring applications on meshes with fabrication errors. Our method enables off-chip configuration of PPIC mesh or can be utilized to furnish initial voltage settings before conducting on-chip configuration.

**Funding.** This work is partly supported by NSFC program (62275029, 61935003, 62021005).

**Disclosures**. The authors declare that there are no conflicts of interest.

**Data availability.** Data underlying the results presented in this paper are not publicly available at this time but may be obtained from the authors upon reasonable request.